\documentclass{aa}
\usepackage{graphicx}
\usepackage{color}
\begin{document}
\title{Spectral variations of the X-ray binary pulsar LMC~X-4 during its
long period intensity variation and a comparison with Her~X-1} 
\subtitle{}

\author{S. Naik and B. Paul}

\offprints{S. Naik : sachi@tifr.res.in}
\institute{Tata Institute of Fundamental Research, Homi Bhabha Road, Mumbai - 400005, India}
\authorrunning{Naik and Paul}
\titlerunning{Spectral Variations in LMC~X-4 and Her~X-1}
\date{Accepted for publication}

\abstract{
We present spectral variations of the binary X-ray 
pulsar LMC~X-4 using the RXTE/PCA observations 
at different phases of its 30.5 day long super-orbital period. 
Only out of eclipse data were used for this study. During the 
high state of the super-orbital period of LMC~X-4, the spectrum is 
well described by a high energy cut-off power-law with a photon 
index in the range of 0.7$-$1.0 and an iron emission line. In 
the low state, the spectrum is found to be flatter 
with power-law photon index in the range 0.5$-$0.7. A direct 
correlation is detected between the continuum flux in 7$-$25 keV 
energy band and the iron emission line flux. The equivalent width 
of the iron emission line is found to be highly variable during 
low intensity state, whereas it remains almost constant during 
the high intensity state of the super-orbital period. It is 
observed that the spectral variations in LMC~X-4 are similar to 
those of Her~X-1 (using RXTE/PCA data). These results suggest 
that the geometry of the region where the iron line is produced
and its visibility with respect to the phase of the super-orbital
period is similar in LMC~X-4 and Her~X-1. A remarkable difference
between these two systems is a highly variable absorption column density
with phase of the super-orbital period that is observed in Her~X-1
but not in LMC~X-4.
\keywords{stars: neutron- Pulsars: individual: (LMC~X-4, Her~X-1) -X-rays: 
stars}
}
\maketitle
\section{Introduction}
LMC~X-4 is an eclipsing high-mass disk-fed accretion-powered binary 
X-ray pulsar in the Large Magellanic Cloud. It was first detected with the
{\it Uhuru} satellite. The optical counterpart of the pulsar is found to be 
a 14th magnitude OB star (Sanduleak \& Philip 1977). Photometric and 
spectroscopic observations of the star revealed strong ellipsoidal light 
variations which indicated a binary period of 1$^d$.408 (Chevalier \& 
Ilovaisky 1977). X-ray pulsations with a spin period of 13.5 s were discovered 
by Kelley et al. (1983) which was later detected in the EXOSAT observations 
of the source even during the quiescent period and several 1 minute 
flares (Pietsch et al. 1985).
X-ray Eclipses with a 1$^d$.4 recurring period were discovered by Li et al.
(1978) and White (1978). The X-ray intensity varies by a factor of $\sim$ 
60 between high and low states with a periodic cycle time of 30.5 day (Lang 
et al. 1981). The 30.5 day intensity variation has also been detected
with the All Sky Monitor (ASM) detectors of GINGA and {\it The Rossi X-ray
Timing Explorer (RXTE)} (Paul \& Kitamoto 2002). Flux modulation 
at super-orbital period in LMC~X-4 is believed to be due to blockage of 
the direct X-ray beam by its precessing tilted accretion disk, as in the 
archetypal system Her~X-1. Flaring events of duration ranging from 
$\sim$ 20 s to 45 minutes (Levine et al. 1991 and references therein) are 
seen about once in a day during which the source intensity increases by 
factors up to $\sim$ 20. 

GINGA and ROSAT observations showed that the X-ray spectrum of LMC~X-4 in 
0.2$-$30 keV energy range can be modeled with a power law component 
with photon index $\alpha$ $\sim$ 0.7, a high energy cutoff E$_c$ $\sim$ 
16 keV and a folding energy E$_F$ $\sim$ 36 keV, a thermal bremsstrahlung 
component with temperature of 0.35 keV and a blackbody with kT $\sim$ 0.03 
keV (Woo et al. 1996).  The spectrum also shows a broad iron emission line 
at E $\sim$ 6.6 keV. The soft excess, whether modeled as a thermal
Bremsstrahlung or a black-body type emission, shows modulation at the
pulse period (Woo et al. 1996 and Paul et al. 2002). Paul et al. (2002)
also argued that the  Bremsstrahlung model of LMC~X-4 is physically not
acceptable due to its pulsating nature. 
The soft excess, also detected with 
BeppoSAX (La Barbera et al. 2001) was modeled as blackbody emission from 
accretion disk at magnetospheric radius Comptonized by moderately hot 
electrons in a corona above the accretion disk. La Barbera et al. (2001) 
also reported the presence of a cyclotron absorption line at $\sim$ 100 
keV in the 0.12$-$100 keV energy spectrum of the source.

\begin{figure}
\vskip 7.7 cm
\includegraphics{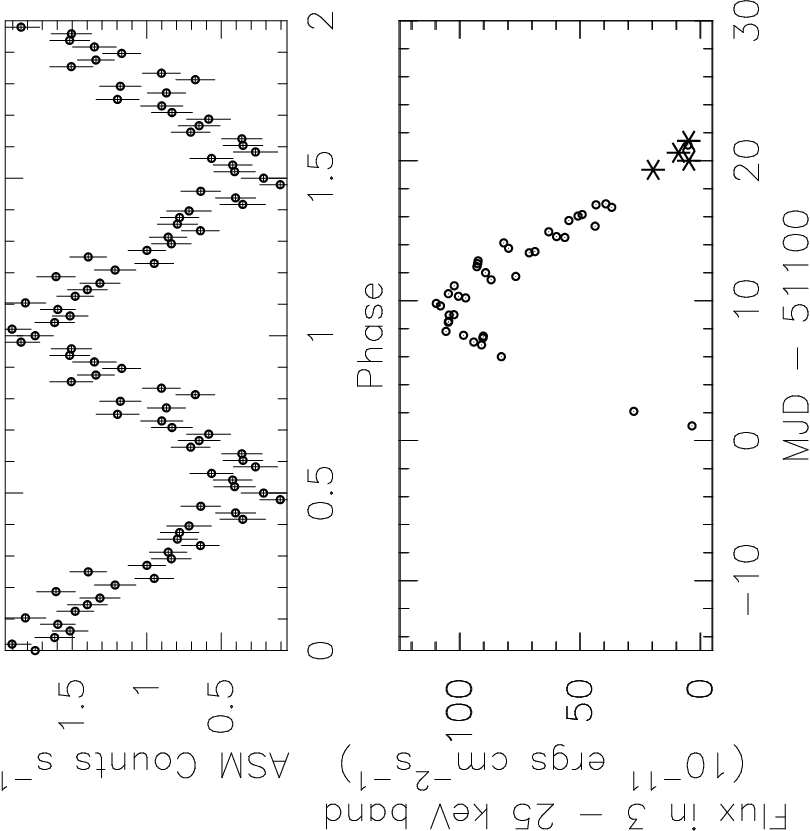}
\caption{The folded light curve of LMC~X-4 in 1.3$-$12.1 keV energy band
over 30.5 day super-orbital period, obtained from RXTE/ASM data is shown 
(in upper panel) along with the background-subtracted X-ray flux (averaged 
over each orbit of the RXTE/PCA observations) in 3$-$25 keV energy range 
(bottom panel) during the 1998 October-November observations. The points 
marked by ``{\LARGE $*$}'' are for the observations which were made
outside the selected time range and have been included here based on the phase
of the super-orbital period. These observations are used to get a better
coverage of low intensity state.}
\end{figure}

The accretion-powered X-ray pulsar Her~X-1 with a spin period of 1.24 s  and 
orbital period of 1.7 days is another binary system which possesses a long-term 
cycle of 35 days in X-ray intensity, analogous to LMC~X-4. The long-term
intensity variation in Her~X-1 is believed to be due to the obscuration by
a precessing tilted accretion disk around the central neutron star counter 
rotating with respect to the orbital motion (Gerend \& Boynton 1976). 
The X-ray light curve of the source over 35 day super-orbital period consists 
of main-on X-ray state with a mean duration of $\sim$ 7 orbital periods which 
is surrounded by two off-states each of $\sim$ 4 orbital cycles and a short-on 
state of smaller intensity with a duration of $\sim$ 5 orbits.
From GINGA observations it was found that the iron line intensity is well
correlated with the total continuum intensity 
whereas the iron line equivalent width varies between 0.48 keV to $\simeq$ 
1.37 keV during the 35 day phase (Leahy 2001). As both the X-ray 
pulsars Her~X-1 and LMC~X-4 show long periods in similar time scale, we 
make a comparison of the spectral properties of LMC~X-4 with the 
well known source Her~X-1.

In this paper we discuss the spectral variations of LMC~X-4 and Her~X-1
during the long periods using archival data from the observations with the 
Proportional Counter Array (PCA) of the RXTE and compare the spectral 
properties of these two sources. 

\begin{figure}
\vskip 7.7cm
\includegraphics{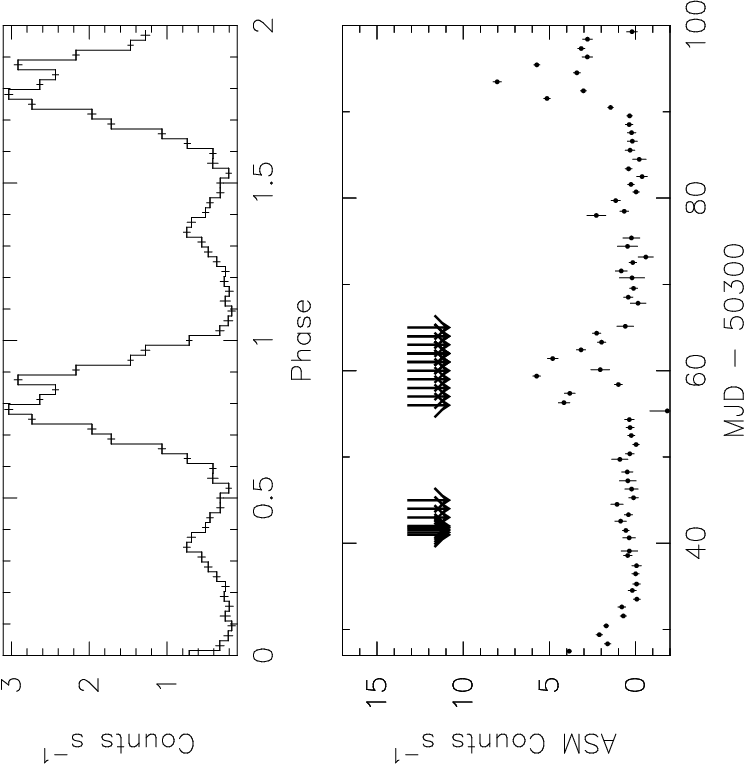}
\caption {The folded light curve of Her~X-1 in 1.3$-$12.1 keV energy band 
over 35 day super-orbital period, obtained from RXTE/ASM data is shown (in upper
panel) along with the RXTE/ASM light curve (one day averaged data) of the 
source for two super-orbital period (bottom panel) which shows the presence 
of main-on, short-on, and low state. The RXTE/PCA observations used for the 
spectral analysis are shown by arrows in the figure.}
\end{figure}

\begin{figure*}
\vskip 10.5 cm
\includegraphics{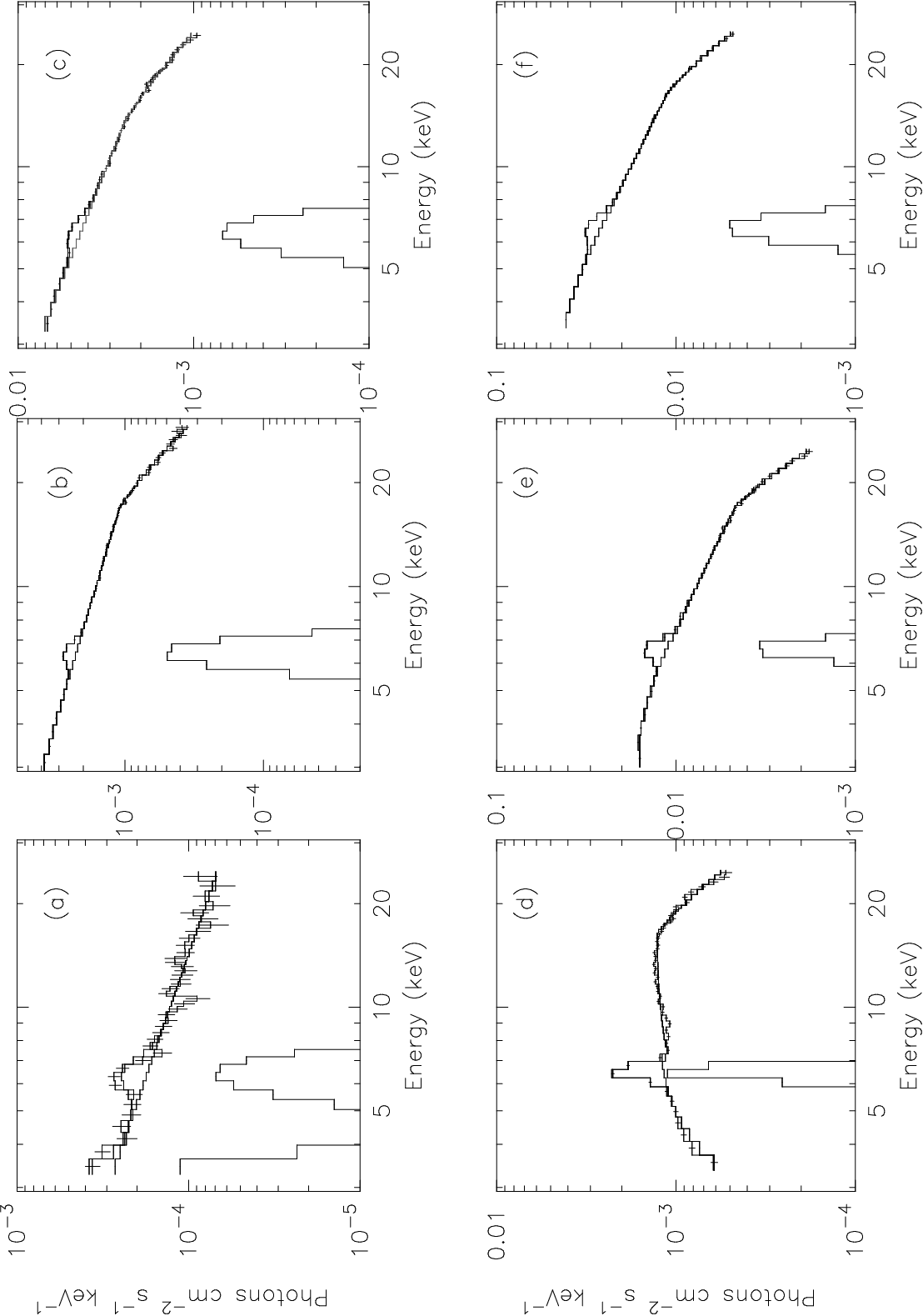}
\caption{The deconvolved X-ray spectra of LMC~X-4 (top panels) and Her~X-1
(bottom panels) during low intensity states (a) and (d), medium intensity states
(b) and (e) which is known as short-on state for Her~X-1 and high intensity 
states (c) and (f) are shown here. The best fit model consists of a 
blackbody, a power-law, a Gaussian function for the iron emission line, and 
a high energy cutoff . }
\end{figure*}

\section{Observations}

To study the super-orbital phase dependence of various spectral parameters,
we have selected RXTE observations of LMC~X-4 and Her~X-1 at different 
phases of the super-orbital periods of 30.5 day and 35 day respectively. For 
LMC~X-4, data from RXTE observations  made mainly between 1998 October 2 
and 1998 November 4 were used for spectral analysis. To have a better coverage 
of low intensity state (phase) during the super-orbital period, we have also 
analyzed data from a few other RXTE observations made on 1998 February 6, 
June 11, August 12, September 9, and December 4. The data used for spectral 
analysis are out-of-eclipse and free from the the flaring state.
Figure 1 shows the folded ASM light curve (upper panel) over 30.5 day 
super-orbital period and the background subtracted source flux in 3$-$25 
keV energy band of all the observations used for spectral analysis in the 
present work. From the figure, it can be seen that the data used for
analysis cover a reasonably good part of the 30.5 day super-orbital period.
We have shown, in Figure 2, the RXTE/ASM light curve of Her~X-1 folded 
at 35 day super-orbital period (upper panel) and the RXTE/ASM light curve of 
the source in time range of two cycles of the 35 day super-orbital period 
(bottom panel). A total number of 17 RXTE/PCA pointed observations of 
Her~X-1, covering the main-on, short-on and low states of the super-orbital 
period are used in the present work.

\begin{figure*}[t]
\vskip 10.5 cm
\includegraphics{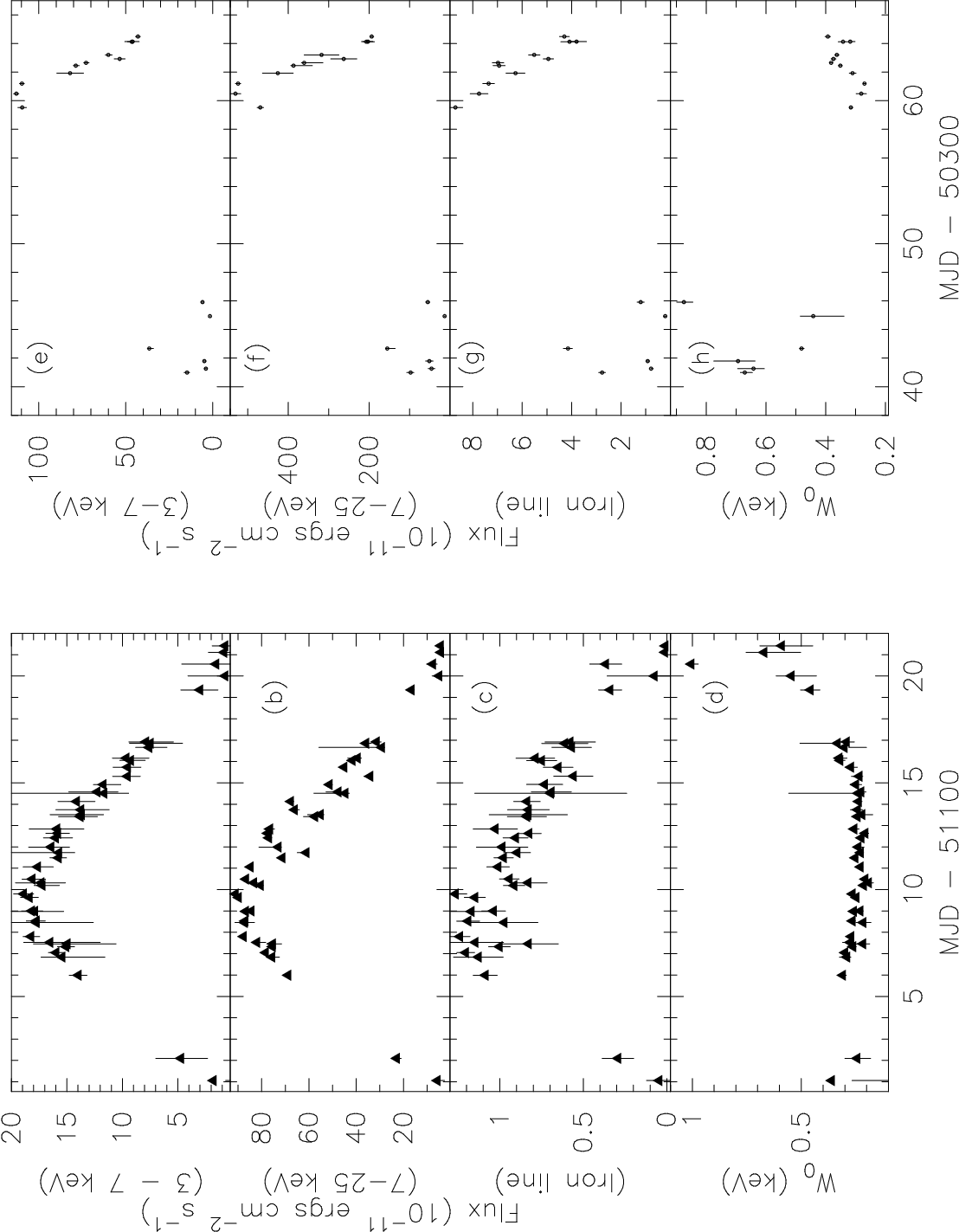}
\caption{The figure shows change in the source flux in 3$-$7 keV, 7$-$25 keV,
iron emission line flux, and iron equivalent width (W$_0$) during the 
super-orbital period of LMC~X-4 (left panels) and Her~X-1 (right panels) 
respectively. The change in the iron emission line flux with the source flux 
in 7$-$25 keV energy band are similar for both the sources over the respective 
super-orbital periods. From the panels (d) and (h), it is seen that the low 
intensity states are characterized by high value of iron equivalent width 
whereas the high intensity states are characterized by low value of iron 
equivalent width which is similar for both the sources.}
\end{figure*}

\begin{figure*}[t]
\vskip 10.5 cm
\includegraphics{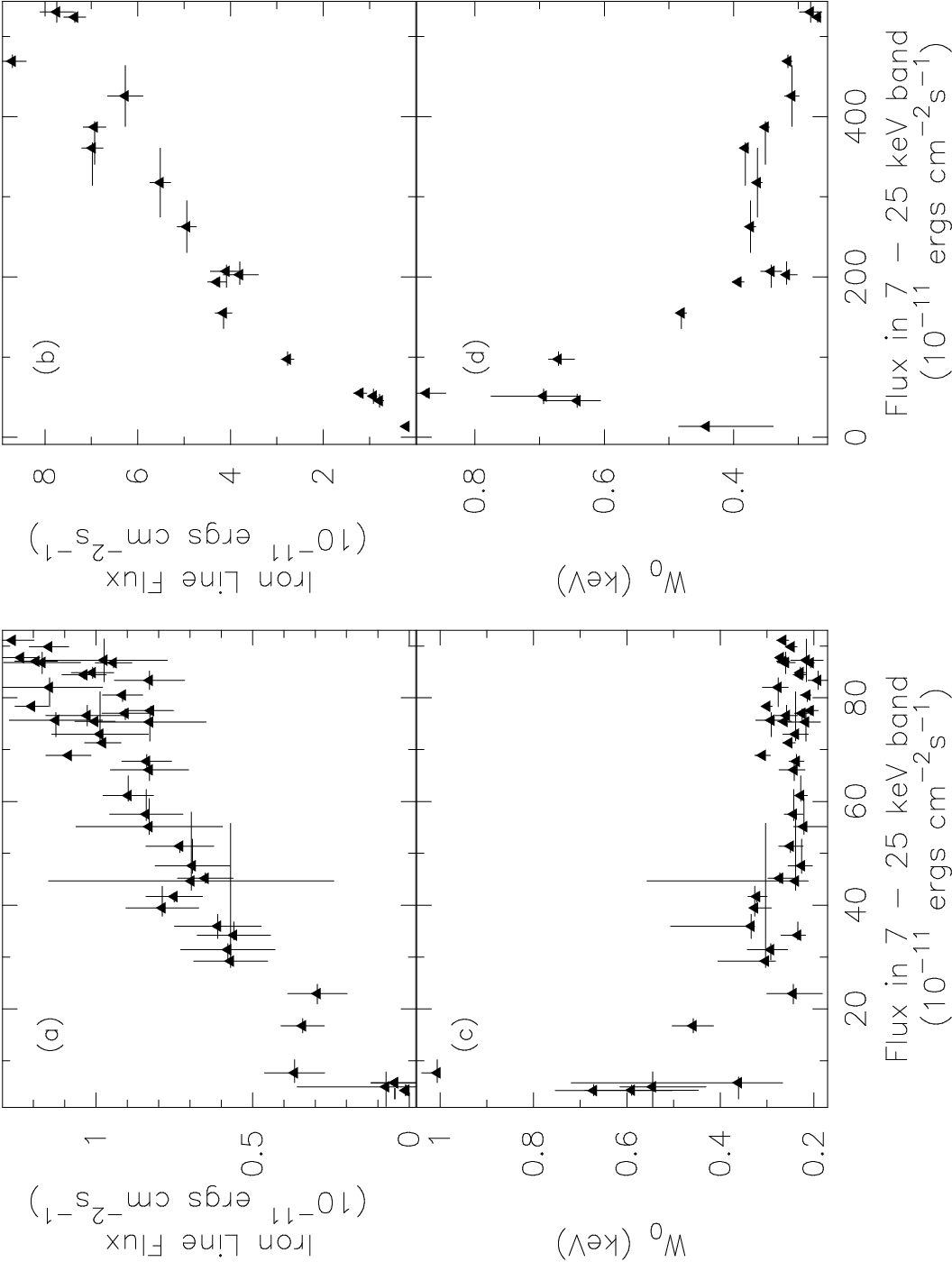}
\caption{The figure shows the change in the iron emission line flux and 
iron equivalent width (W$_0$) with the source flux in 7$-$25 keV energy 
band over the super-orbital period of LMC~X-4 (left panels) and Her~X-1 
(right panels). It is seen that the low intensity states are characterized 
by high value of iron equivalent width whereas the high intensity states 
are characterized by low value of iron equivalent width which is similar 
for both the sources.}
\end{figure*}

\section{Spectral Analysis and Results}

\subsection{Data Reduction}

Energy spectra in 129 channels were generated from the Standard 2
mode PCA data. The 
standard procedures for data selection, and response matrix generation were 
followed. Background estimation was done using both bright and faint models 
of RXTE/PCA according to different intensity states of the source at different 
phases. We restricted our analysis to the 3$-$25 keV energy range. Data from all
the available, at the time of observation, Proportional Counter Units (PCUs)
were added together and response matrices were generated accordingly.

\subsection{Choice of Spectral models}

\subsubsection{LMC~X-4}

In the energy band of 2--25 keV, the energy spectrum of LMC~X-4
is known to consists of a single power law with high energy cutoff, 
a soft excess, iron K shell emission line, and low energy absorption.
The functional form of the model used for spectral fitting is 
\\
\\
\begin{math}
I(E) = exp [-\sigma(E)N_H]
\end{math}
\begin{equation}
~~~~~~~~~\times [f_{PF}(E) + f_{PL}(E) + f_{FE}(E)] \times f_{hi}(E)
\end{equation}
where ~~$f_{PF}(E) = Planck~~ function \\
   = I_{PF}(\frac{E}{E_{PF}}^2 (e-1)[exp(\frac{E}{E_{PF}}) - 1]^{-1}$,\\
$f_{PL}(E) = Power~~ law = I_{PL} E^{-\alpha}$, \\
$f_{FE}(E) = Gaussian-shaped~~ iron~~ emission~~ line \\
= I_{FE} (2\pi\sigma_{FE}^2)^{-1/2} exp [-\frac{(E - E_{FE})^2}{2\sigma_{FE}^2}$], \\ 
$f_{hi}(E) = high~~ energy~~ cut-off \\
$~~~~~~~~~~~~~~~~$ = 1, ~~~~~~~~~~~~~~~~  if~~~ E < E_c$ \\
 = exp ($-\frac{E - E_c}{E_f}), ~~~~ if~~E~~\geq~~E_c$\\

The value of absorption by intervening cold material parameterized as 
equivalent Hydrogen column density N$_H$ was kept fixed at of 0.055 $\times$ 
10$^{22}$ cm$^{-2}$ which is the Galactic column density towards this source. 
The blackbody temperature (kT), the center and the width of the iron emission 
line were fixed at 0.2 keV, 6.4 keV and 0.65 keV respectively with free 
normalization (Naik \& Paul 2002). 

\subsubsection{Her~X-1}

The 2$-$37 keV energy spectrum of Her~X-1 in the main high, short high and
low state was measured with GINGA and two component power-law continuum
(absorbed and unabsorbed) model was found to fit the data well (Leahy 2001).
However, the best broad-band energy spectrum (0.1$-$200 keV) measured
with the Beppo-SAX does not seem to require the double power-law components
and is well described by a model consisting of a single power-law
with a high energy cut-off, a blackbody for the low energy excess, two
Gaussian iron emission features near 1.0 keV and 6.5 keV, and a $\sim$
40 keV cyclotron absorption feature (Dal Fiume et al. 1998).
Oosterbroek et al. (2000, 2001) also used a model, consisting of a soft
black-body and a single hard power-law together with two Gaussian emission 
features near 1.0 keV and 6.5 keV, to fit the 0.1$-$30 keV spectra of 
Her~X-1 from BeppoSAX. We have, therefore, assumed the same spectral 
model as for LMC~X-4.
For Her~X-1, we have kept the blackbody temperature 
fixed at $kT = 0.1$ keV (Vrtilek et al. 1994). However, unlike in the case 
of LMC~X-4, the equivalent Hydrogen column density, center of the iron 
emission line, and width of the iron emission line were not fixed. 

\subsection{Results}

Figure 3 shows the deconvolved X-ray spectra of LMC~X-4 and Her~X-1
during low intensity states (Obs. IDs : 30125-04-11-01 and 10055-01-02-00), 
medium intensity states (Obs. IDs : 30085-01-31-00 and 10055-01-20-00), 
and high states (Obs. IDs : 30085-01-18-00 and 10055-01-13-00)
of the 30.5 day and 35 day super-orbital periods respectively. The presence 
of prominent iron emission lines in the spectra of two sources during the 
low intensity states are clearly seen in the panels (a) and (d) respectively. 
The equivalent width of the iron line is reduced with increase in the intensity 
for both the sources as seen in the panels (b), (c) and (e), (f) respectively. 
Although the RXTE/PCA is not sensitive at low energies ($\leq$ 3 keV), the 
variation in $N_H$ over the 35 day super-orbital period for Her~X-1 is 
very significant (from 1.7 $\times$ 10$^{20}$ cm$^{-2}$ in high state
to 7.3 $\times$ 10$^{22}$ cm$^{-2}$ in low state). The black-body component 
is undetected in both sources except in a few low state spectrum of 
LMC~X-4 where the power-law component is comparatively weak. 

We have estimated the continuum flux in 7$-$25 keV and the iron line flux 
for all the RXTE/PCA observations of LMC~X-4 and Her~X-1 used in the 
present work. The center energy, flux, and equivalent width of the iron
line, and the continuum flux in 7$-$25 keV energy band during one
representative observation in low, medium, and high intensity states of
Her~X-1 and LMC~X-4 are given in Table 1.

The left panels in Figure 4 show the continuum flux  
in 3$-$7 keV and 7$-$25 keV energy range, the iron emission 
line flux, and the equivalent width with respect to phases of the 
super-orbital period of LMC~X-4. The same quantities for
Her~X-1 are shown in the right panels of Figure 4. From this figure, 
it can be seen that the relation between hard X-ray flux and iron line 
flux is remarkably similar in LMC~X-4 and Her~X-1. In Figure 5, we 
show the change in the iron emission line flux and equivalent width  
as a function of the hard X-ray flux (7$-$25 keV)
in LMC~X-4 (left panels) and Her~X-1 (right panels) during the 
30.5 day and 35 day super-orbital periods respectively. From the
upper panels of the Figure 5, a global proportionality between the
line flux and the hard X-ray continuum flux can be seen for both
LMC~X-4 and Her~X-1. Similar type of features have also been 
observed in Vela~X-1 and GX~301-2 (Nagase 1989). Although no 
clear correlation or anti-correlation is seen between the source 
flux and the iron line equivalent width ($W_0$), it is observed 
that $W_0$ of LMC~X-4 is very high (0.4$-$1.1 keV) during low 
intensity states (source flux $\leq$ 2.0 $\times$ 10$^{-10}$ ergs 
cm$^{-2}$ s$^{-1}$). During high intensity states, no significant 
change in $W_0$ is observed and its value lies in 0.2$-$0.4 keV 
energy range for most of the times of the 30.5 day period. The change 
in $W_0$ with the source flux is similar for Her~X-1 during different 
intensity states of the super-orbital period. It is also observed that 
the iron line energy is $\geq$ 6.6 keV during the main-high state of 
the 35 day super-orbital period of Her~X-1 which decreases to 6.4$-$6.55
keV during the low state of the source. These results are consistent 
with the previously reported individual measurements of iron line energy 
from Her~X-1. However, no systematic changes are observed in the 
power-law index, high energy cut-off with the source flux in 7$-$25 keV 
for LMC~X-4 and Her~X-1 during the super-orbital intensity variation. 
Apart from the change in the Hydrogen column density (N$_H$) during the 
super-orbital period of Her~X-1 which is constant in LMC~X-4 (Lang 
et al. 1998), the change in all other spectral parameters are similar 
for both the sources. 

\begin{table*}[t]
\centering
\caption{Spectral fit parameters of LMC~X-4 and Her~X-1 during low, medium
and high intensity states}
\begin{tabular}{llllll}
\hline
\hline
State  &N$_H^1$  &Fe$_c^2$  &W$_0^3$  &\multicolumn{2}{l}{Flux (10$^{-11}$ erg cm$^{-2}$ s$^{-1}$)}\\
     &(10$^{22}$ cm$^{-2}$) &(in keV) &(in keV) &7$-$25 keV &Iron line \\
\hline
\hline
\multicolumn{5}{c}{Her~X-1} \\
\hline
Low   &8.7   &6.49$\pm$0.03  &0.65$^{+0.045}_{-0.036}$ &45.63$\pm$8.26 &3.96$\pm$0.75\\
Medium &3.96 &6.6$\pm$0.04   &0.32$\pm$0.03 &202.9$_{-11.97}^{+15.96}$ &3.8$\pm$0.3  \\ 
High  &1.9   &6.62$\pm$0.02 &0.28$\pm$0.02 &530.6$\pm$12.7  &7.7$\pm$0.35\\
\hline
\multicolumn{5}{c}{LMC~X-4} \\
\hline
Low     &0.055$^*$  &6.4$^*$  &0.67$_{-0.168}^{+0.082}$ &4.3$_{-0.97}^{+1.15}$ &0.12$\pm$0.06\\
Medium  &0.055$^*$  &6.4$^*$  &0.322$\pm$0.022  &41.7$\pm$1 &0.75$\pm$0.09\\
High    &0.055$^*$  &6.4$^*$  &0.26$_{-0.02}^{+0.018}$ &87$\pm$2  &1.17$\pm$0.12\\
\hline
\hline
\multicolumn{6}{l}{N$_H^1$: Equivalent hydrogen column density,~~~~Fe$_c^2$: Iron line energy,}\\
\multicolumn{6}{l}{~~~~~~~~~~~W$_0^3$: Iron equivalent width}\\
\multicolumn{6}{l}{$^*$: Parameters are fixed during the spectral fitting as described in the text.}\\
\end{tabular}
\end{table*}

\section{Discussion}

Iron emission lines in X-ray pulsars are produced by illumination of
neutral or partially ionized material either in (a) accretion disks 
(mostly seen in LMXB pulsars), (b) stellar wind of the high mass companion star,
(c) material in the form of an circumstellar shell, far away from the star, 
(d) material in the line of sight, or (e) the accretion column. Inoue (1985) 
and Makishima (1986) estimated the equivalent widths of the fluorescence iron 
line emission from neutral matter in a sphere surrounding the X-ray source 
using a power law type incident spectrum. In accretion powered X-ray 
pulsars, the iron equivalent width can be higher if the compact object is 
hidden from direct view by the accretion disk and only X-rays scattered into 
the line of sight by an accretion disk corona or wind are visible.

If absorbing matter is located between the X-ray source and the 
observer, the continuum spectrum is absorbed by the matter resulting in
increase in the equivalent width monotonically with the column density.
This effect was observed in GX~301-2, in which a linear correlation
between the equivalent width of iron emission line and the column density
was found and was interpreted as cold isotropic gas surrounding the compact
source (Makino et al. 1985).
From ASCA observations of GX 301-2, Endo et al. (2000) found that the 
iron K$_\alpha$ emission line originates from the region typically at 
around several {Alfv\'en radii} from the neutron star and the observed 
correlation between the equivalent width of iron K$_\alpha$ line and the 
line of sight iron column density is consistent with the picture that the 
fluorescing matter surrounds the pulsar by $\sim$ 4$\pi$ solid angle.
In the HMXB pulsar Vela~X-1, the variation in iron line intensity
is found to be correlated with the continuum flux variation and the
equivalent width of the iron line is found to become relatively large
with increase in in-homogeneity of the surrounding matter
(Nagase et al 1986). As Vela~X-1 is 
powered by accretion from stellar wind, three locations other than 
the accretion disk (as described above) may be responsible for iron line 
emission. GINGA and ASCA observations of the iron emission line in GX~1$+$4
suggest the origin to be an isotropically distributed cold matter 
at a distance of $\sim$ 10$^{12}$ cm (Kotani et al. 1999).

In the previous section we presented detection of a positive correlation
between the iron emission line flux and the hard X-ray continuum flux in
LMC~X-4 and Her~X-1 and variability of the equivalent width between different
intensity states. For Her~X-1, this was already reported by Leahy (2001),
who inferred that reprocessing of X-rays is important during all intensity
levels and the major source of the iron line is the accretion column
which is consistent with the accretion-column inner-disk model
of Scott et al. (2000). The iron line of Her~X-1 is now actually known to
consists of two components at 6.41 keV and 6.73 keV, which indicates that
the observed changes in iron line energies are due to a varying mixture
of the two lines (ASCA; Endo et al. 2000). Probably the 6.7 keV line
originates in a highly ionized accretion column whereas the 6.4 keV
line is associated with the reprocessing in the accretion disk.

The similarities in the variability pattern of iron line flux and
hard X-ray continuum flux of Her~X-1 and LMC~X-4 suggest a similar
iron line origin. The identical changes in the iron line parameters 
over respective super-orbital periods for both the sources reveal the 
similarities in the distribution of matter in the system causing the 
iron line emission. However, we note that the Hydrogen column 
density was found to be almost constant over the super-orbital period of 
LMC~X-4 (Lang et al. 1998) which is highly variable in case of Her~X-1.
This can be explained by the absence of the significant amount of neutral 
matter along the line of sight. Heindl et al. (1999) explained that 
in LMC~X-4, the flux modulation is caused by the accretion disk and it 
occurs at the hot and ionized inner disk region where Thomson scattering 
can remove photons from the beam without affecting the overall spectral shape. 

In both the pulsars, the correlation between iron line flux and total flux (7$-$25 keV)
in high state indicates that the iron line is produced near
the continuum X-ray source and both are obscured by a precessing 
accretion disk. However, when the obscuration is very high i.e, in 
the low state, the equivalent width increases. This indicates that 
there may be a second, weaker source of iron line away from the hard 
X-ray source which is not strongly related to the super-orbital 
intensity variation. Super-orbital phase resolved observations
with good low energy response can help to solve the issue of
column density variation.

\section{Acknowledgment}
We express sincere thanks to the referee for his useful
comments and suggestions which improved the contents and presentations
of the paper.
The work of SN is partially supported by the Kanwal Rekhi Scholarship of
the TIFR Endowment Fund. This research has made use of data obtained 
through the High Energy Astrophysics Science Archive Research Center 
Online Service, provided by the NASA/Goddard Space Flight Center.

\end{document}